\documentstyle[fleqn,twoside,gc-99]{article}

\heads{P.S. Negi and M.C. Durgapal}
     {Status of a Self-Bound Equations of State and Analytic
	Solutions}

\begin{document}
\twocolumn[
\Arthead{7}{2001}{1 (25)}{1}{4}

\Title{STATUS OF A SELF-BOUND EQUATIONS OF STATE  \yy
       AND ANALYTIC SOLUTIONS IN GENERAL RELATIVITY}

\Author{P.S. Negi\foom 1 and M.C. Durgapal\foom 2}
{Department of Physics, Kumaun University, Nainital-263 002, India}

\Rec{23 December 1999}

\Abstract
{We have obtained a criterion for spherically symmetric and static
structures under hydrostatic equilibrium in general relativity (GR), which
states that for a given value of $\sigma\ \equiv (P_0/E_0) \equiv $ the
ratio of central pressure to  central energy-density], the compaction
parameter $u \equiv (M/a)$, where $M$ is the total mass and $a$ is the
radius of the configuration] or the surface redshift of any regular
configuration cannot exceed that of the corresponding homogeneous density
sphere, that is, $u \leq u_h$, where $u_h$ is the compaction parameter of
the homogeneous density sphere. By examining various exact solutions and
equations of state available in the literature, we find that this criterion
is fulfilled only by those configurations in which the surface density
vanishes alongwith the pressure (meaning thereby that the pressure, energy
density, metric parameters and their first derivatives are continuous on the
surface). On the other hand, configurations having a finite density on
the surface (that is, the self-bound structures) do not fulfill this
criterion. This criterion puts a severe restriction on the static structures
based upon the general relativistic field equations and consequently on the
upper limit of mass, surface and central redshift and other physical
parameters of spherically symmetric and static configurations.}

\vspace{58mm}
%

] 
\email 1 {psnegi\_nainital@yahoo.com}

\section{Introduction}

The exact solution for an incompressible fluid sphere of uniform energy
density, $E$, in GR was first obtained by Schwarzschild [1]. In spite of its
nonphysical behaviour (the adiabatic speed of sound $v = \sqrt{dP/dE}$ is
infinite throughout the structure, and the pressure vanishes at finite
surface density), it has important features characterizing configurations
compatible with the structure of general relativity: (i) it gives an
absolute upper limit for the compaction parameter, $u [\equiv (M/a)$, mass
to size ratio of the entire configuration in geometrized units] $ \leq
(4/9)$ for any regular static solution in hydrostatic equilibrium [2], and
(ii) for an assigned value of the compaction parameter $u$ and the radius $a$, the minimum
central pressure $ P_0 $ corresponds to the homogeneous density solution
(see, e.g., [3]).

In this paper, by the use of important feature (ii) mentioned above, we extract
the following statement : for a given value of $\sigma [\equiv (P_0/E_0)$ ratio
of the central pressure and central density] the maximum value of
the compaction parameter corresponds to the homogeneous density sphere.

Chandrasekhar [4] discussed the hydrostatic equilibrium condition under
small adiabatic perturbations and showed that for a {\em compressible\/}
homogeneous density distribution, as long as the ratio of the specific
heats $\gamma$ remains finite, a dynamical instability intervenes before
the Schwarzschild limit $(u = 4/9)$ is reached. And for $ \gamma \to
\infty,\ u \to (4/9)$. Thus for any physically viable solution one
may expect a {\em finite\/} value of $\gamma$, and the dynamical instability
would intervene well before the compaction parameter $u$ approaches the
limiting value of (4/9).

Exploring various physical properties of a superdense object like a neutron
star, one may expect to have some physically viable equation of state (EOS).
However, for such objects the equations of state are not well known
(empirically) because of the lack of knowledge of nuclear interactions
beyond the density $ \cong 10^{14}$ g/cm$^{3}$ [5], and the only way of
obtaining an EOS far beyond this density range is extrapolation. Various
such extrapolated equations are available in the literature [6]. As a way
out, one can impose some restrictions upon the known physical quantities,
such that the speed of sound inside the configuration, $v [\sqrt (dP/dE)]$,
does not exceed the speed of light in vacuo, i.e., $v \leq c = 1$ (in
geometrized units), and obtain an upper bound on stable neutron star masses
[7--9].  Haensel and Zdunik [10] have shown that the only EOS which can
describe a submillisecond pulsar and the static mass of $1.442 M_\odot $
simultaneously, corresponds to the EOS $(dP/dE) = 1$, however, they
emphasized that this EOS represents an ``abnormal'' state of matter in the
sense that the pressure vanishes at densities of the order of nuclear
density or even higher.

Alternatively, one may assume an expression for one of the parameters,
$ P, E, \nu, $and$ \lambda$ in terms of the radial coordinate $r$
(or an algebraic relation between the parameters) and obtain an exact
solution of the Einstein field equations. Many exact solutions are
available in the literature [11--16], which may be used to obtain various
physical properties of spherical and static compact object (provided they
are physically realistic).

Various EOS and exact solutions discussed in the literature, in fact,
fulfill the criterion (i), that is, the equilibrium configurations
pertaining to these EOS and analytic solutions correspond to the compaction
parameter $u$ which is always smaller than the value 4/9. However, we show
in the present paper that the EOS or analytic solutions, corresponding to a
{\em finite\/} surface density (where the pressure vanishes at finite
surface density, the so-called ``self-bound'' state of matter), in fact,
are not compatible with the structure of GR, as they do
not fulfil the criterion (ii) (namely, for an assigned value of $\sigma$
the compaction parameter $u$ of any regular configuration should not be
greater than that of the homogeneous density configuration). We have shown
this inconsistency particularly for the EOS $(dP/dE) = 1$ (as it represents
the most successful EOS for obtaining various extreme characteristics of
neutron stars), and for the exact solution put forward by Durgapal and
Fuloria [15] (as it represents various characteristics expected for a
physically realistic superdense object [17]). Furthermore, it can be shown
that this inconsistency always exists for any EOS or analytic solution with
finite surface density. On the other hand, we have found that only those EOS
or exact solutions which correspond to vanishing density at the surface
of the configuration (i.e., gravitationally bound structures) are
compatible with the structure of GR. We show that the
above criterion is fulfilled for polytropic equations of state, $P =
KE^\Gamma $ (where $K$ and $\Gamma$ are constants [18]), and $P =
K\rho^{\Gamma_1}$ (where $K$ and $\Gamma_1$ are constants and $\rho$ is the
rest mass density [19]), and Tolman's type VII exact solution [20], $E =
E_0 (1 - r^2 /a^2)$, where $E_0$ is the central energy density and $a$
is the radius of the configuration.

Thus, an important consequence of the EOS or analytic solutions with
vanishing surface density lies in the fact that the ``abnormality'' in the
sense discussed in the literature disappears. Or, in other words, {\em the
EOS or analytic solutions, compatible with the structure of general
relativity, would always represent configurations corresponding to the
normal state of matter\/}.

\section {Criteria for static spherical configurations to be
consistent with the structure of GR}

Let us consider homogeneous sphere of uniform energy-density $E$. The
equations for isotropic pressure $P$ and density $E$ can be written in terms
of the compaction parameter $u$ and the radial coordinate measured in
units of configuration size, $y \equiv r/a$, as
\bear
	8\pi E a^2 \eql 6u,
\\
	8\pi P a^2 \eql 6u \frac{(1 - 2uy^2)^{1/2} - (1 -2u)^{1/2}}
    	 		{3(1 - 2u)^{1/2} - (1 - 2uy^2)^{1/2}}.
\ear

Consider a {\em regular\/} variable density sphere (with some given EOS or
analytic solution) with central energy density $E_0$ and central pressure
$P_0$ , corresponding to the compaction parameter $u = u_v$.

Now, we can always construct a homogeneous density sphere with the same
value of the compaction parameter $u_v$ and energy-density $E_0$, because if
$P_{0h}$ corresponds to the central pressure of this sphere, the ratio
$\sigma_h (\equiv P_{0h} /E_0)$ depends only on the assigned value of the
compaction parameter $u_v$ . And, $P_{0h}$ is given by
\beq
	P_{0h} = \frac{6u}{8\pi a^2}\
		 \frac {1 - (1 - 2u)^{1/2} } {3(1 -2u)^{1/2} - 1} .
\eeq
Now, according to the feature (ii) [3], we may write
\beq
	P_0 \geq P_{0h}, 		
\eeq
or
\beq
	P_0 /E_0 \geq P_{0h}/E_0.
\eeq
Hence for a given value of $u\equiv u_v$, we obtain
\beq
	\sigma_v \geq \sigma_h
\eeq
where $\sigma_v$ is defined as the ratio $ P_0 /E_0$.

\def\uh{u_{\rm h}}
\def\stff{_{\rm stff}}
\def\rptr{_{\rm rptr}}
\def\cptr{_{\rm cptr}}

\begin{table*}

\noi
{\bf Table 1.} \ Various values of the compaction parameter
$u(\equiv M/a)$ corresponding to the exact solutions:
$E = E_0(1 - r^2/a^2)$ (indicated by $u_{\rm tvii}$) and
$E = [8(9 + 2x + x^2)/7(1 + x)^3]$ (indicated as $u_{\rm dfn}$), and
equations of state: $P = KE^\Gamma $ for $\Gamma = 2$ (indicated as
$u\rptr$), $P = K\rho^{\Gamma_1}$ for $\Gamma_1 = 2$ (indicated as
$u\cptr$) and $P = E - E_s$ (indicated by $u\stff$, for different assigned
values of $\sigma \equiv P_0/E_0$. The compaction parameter of a
homogeneous density distribution (Schwarzschild's interior solution) is
indicated as $\uh$ for the same values of $\sigma$. It is seen that
$u_{\rm tvii}, u\rptr$ and $u\cptr \ < u_h$, while $u_{\rm dfn}$ and
$u\stff \ > u_h$. Thus configurations with vanishing surface density
show compatibility with the structure of GR, while those
with finite surface density do not show this compatibility.

\medskip\
\centering

\begin{tabular}{ccccccc}

\hline
$\sigma\equiv P_0 / E_0$ & $u\rptr$    & $u\cptr$  &
	$u_{\rm tvii}$  & $\uh$   &  $u_{\rm dfn}$   & $u\stff$  \\
\hline

0.1252  & 0.1565 & 0.1559 & 0.1588 & 0.1654 & 0.1718 & 0.1683   \\

0.1859  & 0.1950 & 0.1926 & 0.1992 & 0.2102 & 0.2187 & 0.2150   \\

0.2202  & 0.2109 & 0.2076 & 0.2166 & 0.2301 & 0.2392 & 0.2354   \\

0.2800  & 0.2331 & 0.2265 & 0.2407 & 0.2580 & 0.2676 & 0.2646   \\

0.3150  & 0.2434 & 0.2346 & 0.2521 & 0.2714 & 0.2809 & 0.2793   \\

(1/3)   & 0.2481 & 0.2379 & 0.2574 & 0.2778 & 0.2872 & 0.2858   \\

0.3774  & 0.2581 & 0.2441 & 0.2687 & 0.2914 & 0.3003 & 0.3000   \\

0.4350  & 0.2683 & 0.2484 & 0.2809 & 0.3062 & 0.3145 & 0.3153   \\

0.4889  & 0.2765 & 0.2489 & 0.2904 & 0.3178 & 0.3253 & 0.3271   \\

0.5499  & 0.2835 & 0.2466 & 0.2993 & 0.3289 & 0.3354 & 0.3383   \\

0.6338  & 0.2910 & 0.2375 & 0.3092 & 0.3415 & 0.3465 & 0.3501   \\

0.6830  & 0.2945 & 0.2292 & 0.3140 & 0.3476 & 0.3519 & 0.3554   \\

0.7044  & 0.2960 & 0.2250 & 0.3160 & 0.3501 & 0.3541 & 0.3573   \\

0.7085  & 0.2958 & 0.2241 & 0.3164 & 0.3506 & 0.3545 & 0.3577   \\

0.7571  & 0.2985 & 0.2132 & 0.3204 & 0.3558 & 0.3589 & 0.3611   \\

0.8000  & 0.3006 & 0.2033 & 0.3235 & 0.3599 & 0.3624 & 0.3631   \\

0.8360  & 0.3024 & 0.1959 & 0.3260 & 0.3630 & 0.3650 & 0.3638   \\

\hline

\end{tabular}
\end{table*}

Now, varying the compaction parameter $ u_v $ for the homogeneous
density sphere from $u_v$ to $u_h$ (say) we should have
\beq
	\sigma_v = \sigma_h.         
\eeq
For $u = u_h$, the value of $\sigma_h$ would become
\beq
	\sigma_h = \frac {(1 - 2u_h)^{1/2} - 1} {1 - 3(1 - 2u_h)^{1/2} }.
\eeq
Substituting \eq (8) with the help of \eq (7) into (6), we get
\bearr
	\frac {(1 - 2u_h)^{1/2} - 1} {1 - 3(1 - 2u_h)^{1/2} } \geq
        \frac {(1 - 2u_v)^{1/2} - 1} {1 - 3(1 - 2u_v)^{1/2} } . \nnn
\ear
It is clear from \eq (9) that
\beq
		u_h \geq u_v.           
\eeq
That is, for an assigned value of the ratio of central pressure to central
energy-density $\sigma (\equiv \sigma_v)$, the compaction parameter of
homogeneous density distribution $u(\equiv u_h)$ should always be larger
than or equal to the compaction parameter $u(\equiv u_v)$ of {\em any
regular } solution, compatible with the structure of GR. Or,
in other words, for an assigned value of $u$ the {\em minimum\/} value of
$\sigma$ corresponds to the homogeneous density sphere.

\section{Examination of the criterion with some well-known equations
	 of state and exact solutions}

We have considered the EOS $P = (E - E_s)$ (where $E_s$ is the surface
density of the configuration), $P = KE^{\Gamma}$ (where $K$ and $\Gamma$ are
constants), and $P = K\rho^{\Gamma_1}$ (where $K$ and $\Gamma_1$ are
constants and $\rho$ is the rest-mass density). The former EOS
pertains to nonvanishing the latter to vanishing surface density. Similarly,
we have chosen the exact solutions
\[
	(8\pi E/C) = 8(9 + 2x + x^2)/7(1 + x)^3
\]
(where $C$ is a constant and $x = Cr^2$, which corresponds to a
nonvanishing surface density), and
\[
	E = E_0 (1 - r^2 /a^2)
\]
(where $E_0$ is the central energy density and $a$ is the size of the
configuration; for $r = a$, the surface density becomes zero).

Let us denote the compaction parameter corresponding to a homogeneous
density configuration by $\uh$ and that for the equations of state
$P = E - E_s$, $P = KE^\Gamma$ and $P = K\rho^{\Gamma_1}$ by $u\stff$,
$u\rptr$ and $u\cptr$, respectively. For the exact solution
corresponding to the density distribution $(8\pi E/C) = [8(9 + 2x +
x^2)/7(1 + x)^3]$ the compaction is denoted by $u_{\rm dfn}$,
and for $E = E_0 (1 - r^2 /a^2 )$ by $u_{\rm tvii}$, respectively. Now,
for these equations of state and analytic solutions for various
assigned values of $\sigma \equiv (P_0 /E_0)$ we obtain the corresponding
values of the compaction parameters as shown in Table 1. It is seen that
for every assigned value of $\sigma$, $u\rptr$ and $u\cptr \leq \uh \leq
u\stff,$ and $u_{tvii} \leq \uh \leq u_{dfn}$.  Thus we conclude that the
configurations defined by $u\stff$ and $u_{dfn}$ do not show compatibility
with the structure of GR, while those defined by $u\rptr$,
$u\cptr$ and $u_{tvii}$ show such a compatibility. However, this type of
characteristics (the value of the compaction parameter larger or
smaller than $\uh$ for some or all assigned values $\sigma$) can be seen
for any EOS or exact solution having a finite surface density.  On the
other hand, it is seen that the compaction parameter value for EOS or
analytic solution with vanishing surface density always remains smaller
than $\uh$ for all the assigned values of $\sigma$. Therefore we conclude
that to have an EOS or exact solution compatible with the structure of
GR, it is necessary to assure the continuity of density
(and the respective derivative of the metric parameter $\lambda'$)
along with other parameters at the surface of the configuration.

\section{Results and conclusions}

We have investigated various exact solutions and equations of state in
hydrostatic equilibrium and obtained a criterion for the compatibility with
the structure of GR in a physical sense which may be written as: {\em for
an assigned value of the ratio of central pressure to central energy densiy
$\sigma$ the compaction parameter $u$ of any regular solution should not
exceed the compaction parameter $\uh$ of the homogeneous density
distribution.}\ This criterion should be fulfilled by any exact solution
or equation of state, or core-envelope model, or core-mantle-envelope
model, or any complicated distribution of matter, in order to have
compatibility with the structure of the field equations.

Since this criterion imposes a severe restriction on static structures in
GR, namely, to solutions of the Einstein equations corresponding to a
vanishing density at the surface of the configuration, there must appear
new restrictions on the upper limit on mass, surface redshift, central
redshift and other physical properties of static, spherically symmetric
configurations.

\Acknow{The authors acknowledge Uttar Pradesh State
Observatory, Nainital for providing library facilities.}

\end{document}